\documentclass{aa}
\usepackage{natbib}
\bibpunct{(}{)}{;}{a}{}{,}
\graphicspath{{./}{figures/}}

\newcommand\swx{{\it Swift-{\it XRT}}}

\newcommand\cha{{\tt Chandra}}
\newcommand\XMM{{XMM-{\it Newton}}}
\newcommand\NuSTAR{{\tt NuSTAR}}
\newcommand\XSPEC{{\tt XSPEC}}

\newcommand\MYTorus{{\tt MYTorus}}
\newcommand\borus{{\tt borus02}}

\newcommand\uxclumpy{{\tt UXCLUMPY}}
\newcommand\xclumpy{{\tt XClumpy}}
\newcommand\suz{{\tt Suzaku}}
\newcommand\xci{{\tt XCIGALE}}
\newcommand\ci{{\tt CIGALE}}
\newcommand\sfhd{{\tt sfhdelayed}}
\newcommand\bc{{\tt bc03}}
\newcommand\starburst{{\tt dustatt\_modified\_starburst}}
\newcommand\dl{{\tt dl2014}}
\newcommand\skirtor{{\tt skirtor2016}}

\usepackage{courier}
\usepackage[T1]{fontenc}
\usepackage{mathtools}
\usepackage{changepage} 
\usepackage{url,multirow}
\usepackage{appendix}
\usepackage{microtype}
\DeclareUnicodeCharacter{2212}{-}
\usepackage{lineno}
\usepackage{textcomp}

\usepackage[unicode=true,pdfusetitle,
 bookmarks=true,bookmarksnumbered=false,bookmarksopen=false,
 breaklinks=false,pdfborder={0 0 0},pdfborderstyle={},backref=false,colorlinks=true]
 {hyperref}
\hypersetup{
 urlcolor=blue,citecolor=blue}

\usepackage{xcolor}
\usepackage{scrextend}
\usepackage{amsmath}
\usepackage{algorithm}
\usepackage{amsfonts}
\usepackage{mathrsfs}
\usepackage{bm}
\usepackage{rotating}
\usepackage{color}
\usepackage{graphicx}
\usepackage{subfigure}
\usepackage{float}
\usepackage{gensymb}

\makeatletter
\renewcommand*\aa@pageof{, page \thepage{} of \pageref*{LastPage}}
\makeatother

\begin{document}


    \title{A Multi-Wavelength Characterization of the Obscuring Medium at the Center of NGC 6300}



    \author{D. Sengupta\inst{1,2,3}
            \and N. Torres-Albà\inst{4,5}
            \and A. Pizzetti\inst{6}
            \and I.~E. López\inst{1,2}
            \and S. Marchesi\inst{1,2,4}
            \and C. Vignali\inst{1,2}
            \and L. Barchiesi\inst{8,9,10}
            \and I. Cox\inst{4}
            \and M. Gaspari\inst{7}
            \and X. Zhao\inst{11}
            \and M. Ajello\inst{4}
            \and F. Esposito\inst{1,2}
            }
    \institute{Dipartimento di Fisica e Astronomia (DIFA), Università di Bologna, via Gobetti 93/2, I-40129 Bologna, Italy
               \and INAF-Osservatorio di Astrofisica e Scienza dello Spazio (OAS), via Gobetti 93/3, I-40129 Bologna, Italy
               \and Department of Physics, University of Maryland Baltimore County, 1000 Hilltop Circle Baltimore, MD 21250, USA14
               \and Department of Physics and Astronomy, Clemson University, Kinard Lab of Physics, Clemson, SC 29634, USA
               \and Department of Astronomy, University of Virginia, P.O. Box 400325, Charlottesville, VA 22904, USA
               \and European Southern Observatory, Alonso de Cordova 3107, Vitacura, Santiago, Chile
               \and Department of Physics, Informatics and Mathematics, University of Modena and Reggio Emilia, 41125 Modena, Italy
               \and Department of Astronomy, University of Cape Town, Private Bag X3, Rondebosch 7701, Cape Town, South Africa
               \and Inter-university Institute for Data Intensive Astronomy, Department of Astronomy, University of Cape Town, 7701 Rondebosch, Cape Town, South Africa
               \and INAF, Instituto di Radioastronomia-Italian ARC, Via Piero Gobetti 101, I-40129 Bologna, Italy
               \and Department of Astronomy, University of Illinois at Urbana-Champaign, Urbana, IL 61801, USA
               }
               
\titlerunning{NGC 6300}
\authorrunning{Sengupta et al.}

\abstract{Most of the super-massive black holes in the Universe accrete material in an obscured phase. While it is commonly accepted that the ``dusty torus'' is responsible for the nuclear obscuration, its geometrical, physical, and chemical properties are far from being properly understood. In this paper, we take advantage of the multiple X-ray observations taken between 2007 and 2020, as well as of optical to far infra-red (FIR) observations of NGC 6300, a nearby ($z=0.0037$) Seyfert 2 galaxy. The goal of this project is to study the nuclear emission and the properties of the obscuring medium, through a multi-wavelength study conducted from X-ray to IR. We perform a simultaneous X-ray spectral fitting and optical-FIR spectral energy distribution (SED) fitting to investigate the obscuring torus. For the X-ray spectral fitting, physically motivated torus models, such as \borus\/, \uxclumpy\/ and \xclumpy\/ are used. The SED fitting is done using \xci\/. Through joint analysis, we constrain the physical parameters of the torus and the emission properties of the accreting supermassive black hole. Through X-ray observations taken in the last 13 years, we have not found any significant line-of-sight column density variability for this source, but observed the X-ray flux dropping $\sim40-50\%$ in 2020 with respect to previous observations. The \uxclumpy\ model predicts the presence of an inner ring of Compton-thick gaseous medium, responsible for the reflection dominated spectra above 10 keV. Through multi-wavelength SED fitting, we measure an Eddington accretion rate $\lambda_{\rm{Edd}}\sim2\times10^{-3}$, which falls in the range of the radiatively inefficient accretion solutions.}


\maketitle

\section{Introduction}\label{sec:intro}

 According to the unification theory of active galactic nuclei \citep[AGN,][]{urry1995unified}, the accreting supermassive black holes (SMBH) are surrounded by an obscuring medium of dust and gas, commonly referred as `torus'. This torus is homogeneous and obscures the broad line region (BLR) from the line-of-sight (LOS). The torus acts as an absorber of optical-ultraviolet (optical-UV) radiation from the accretion disk of SMBH, which it re-emits at infrared (IR) wavelengths (\citealt{netzer2015revisiting,Ramos_Almeida2017}). However, recent IR observations and analyses of spectral energy distributions (SED) suggest an alternative scenario where the torus exhibits a clumpy structure instead of being homogeneous (e.g., \citealt{nenkova2002dust,Garcia_burillo2021,Ramos_2014}). This is supported by the hydrodynamical simulations of AGN feeding involving Chaotic Cold Accretion (CCA; see \citealt{Gaspari2020} for a review), showing that the AGN environment is continuously shaped by meso ($\sim$kpc) or micro-pc scale cooling clouds that rain from the macro ($\sim$Mpc) scale galactic halos. The variability in LOS obscuration of local AGN in the X-ray spectra supports the clumpy torus scenario (e.g., \citealt{risaliti2002ubiquitous}). LOS X-ray Variability (of hydrogen column density- N$\mathrm{_{H,LOS}}$) due to the obscuration has been identified across a broad range of timescales from approximately one day (e.g., \citealt{Elvis2004,Risaliti2004}) to years (e.g., \citealt{Markowitz2014}). Also, there is a diverse range of observed density changes in obscuration, which are also commonly expected in a CCA precipitation scenario (\citealt{Gaspari2017}): from minor variations of $\Delta(\mathrm{N_{H,LOS}}) \sim 10^{22}$ cm$^{-2}$ e.g., \cite{Laha2020} to the intriguing cases of changing-look AGN, which go from a Compton-thin ($10^{22}$ cm$^{-2}$ < N$_\mathrm{{H,LOS}} < 10^{24}$ cm$^{-2}$) states to Compton-thick (N$_\mathrm{{H,LOS}} > 10^{24}$ cm$^{-2}$) states (e.g., \citealt{risaliti2005rapid,Bianchi2009_CT2CTn,Rivers2015_CT2CTn,Marchesi2022,Serafinelli2019,Mehdipour2023} and more).

Studies with fairly large source samples and regular observations can provide valuable insights into the torus structure. The $\Delta(\mathrm{N_{H,los}})$ method, applied between two observations separated by $\Delta$t, establishes upper/lower limits to cloud sizes and distances to the SMBH (\citealt{risaliti2002ubiquitous,Pizzetti2022,Marchesi2022,Torres-alba_2023}). Along with $\Delta(\mathrm{N_{H,los}})$, we can also study the fraction of flux variability ($\Delta(\mathrm{flux})$) that is not linked with the column density changes, but with a variation in the intrinsic radiation coming from the central engine of the AGN.

This paper is focused on studying the local Seyfert 2 galaxy NGC 6300 ($z$ = 0.0037; RA=$17\degree16'59.47''$, Dec=$-62\degree49'14.0''$). This source is selected from the Compton-Thin sample of \cite{Zhao2021_NustarAGN}, in continuation with the work of \cite{Torres-alba_2023} and \cite{Pizzetti2024}, to investigate the column density variability of Compton-thin AGN in the local Universe (z < 0.1). NGC 6300 is classified as a barred spiral SBb-type galaxy. It has been observed by the Rossi X-ray Timing Explorer (\textit{RXTE}) in 02/1997 (\citealt{Leighly1999}), \textit{BeppoSAX} in 08/1999 (\citealt{Guainazzi2002}) and \XMM\ in 03/2001 (\citealt{Matsumoto2004}). From these early studies, it was classified as a `transient' or changing-look AGN candidate undergoing through a period of low activity. Later, in five epochs from 2007 to 2016, it was observed nine times using \textit{Chandra X-ray Observatory} (\cha\/), \suz\/ and the \textit{Nuclear Spectroscopic Telescope Array} mission (\NuSTAR\/; \citealt{Harrison2013Nustar}). In \cite{Jana2020}, all these observations were studied through time analysis and X-ray spectral analysis using phenomenological models like \textit{powerlaw, compTT, pexrav} and one of the first physically motivated homogeneous torus models: \MYTorus\ (\citealt{murphy2009x}). They inferred the presence of a clumpy torus using the decoupled configuration (where direct powerlaw, reflected and line components are untied) of \MYTorus\/. The column density derived from the reflection medium (presumed to be the average column density of the torus) was found to be different from the LOS column density. However, they did not find any significant flux or LOS column density variability. They also showed the intrinsic luminosity of the source varied ($\Delta \rm{L_{int}} \sim 0.54 \times 10^{42}$ erg s$^{-1}$) from 2009 to 2016. 

In this work, we have carried out a comprehensive and systematic X-ray spectroscopic analysis of NGC 6300, including a new \cha\ observation taken in 2020. For a better characterization of the obscuring torus, we also used optical-IR SED fitting. Firstly, we conducted the X-ray spectral analysis combining sensitive E$<$10 keV observations by \cha\ and \suz\/, with \NuSTAR\/ data at E$>$3 keV: these observations cover a time period from 2007 to 2020. We examined the torus properties, such as inclination angle, covering factor and column density from an X-ray point of view. This was done by using the latest physical motivated X-ray torus models like \borus\ \citep{balokovic2018new}, \uxclumpy\ \citep{buchner2019x} and \xclumpy\ \citep{Tanimoto2019_Xclumpy} which allow us for a proper geometrical characterization of the obscuring material in both smooth and clumpy configurations. Secondly, using aperture photometry, we extracted fluxes from the optical to far-infrared (FIR) band. Using the fluxes and the output parameters of X-ray spectral fitting, we used the broad-band SED fitting tool \xci\ (\citealt{Yang2020_xcigale}) to infer the torus geometry with its host galaxy properties in the mid-IR and X-rays, taking into account all of the physical processes and components of AGN \citep{Esparza-Arredondo2019, Esparza-Arredondo2021, buchner2019x}. Thus, a joint analysis has been carried out by combining the mid-IR SED-derived view of the obscuring medium with that from X-rays. Along with these two approaches, we have also implemented the procedures of \cite{Marchesi2022,Torres-alba_2023}, using the multi-epoch X-ray monitoring to link flux and hydrogen column density variability in different epochs, revealing the dynamical properties of the obscuring medium.


The data reduction processes from X-ray observations and optical-FIR photometry selection procedures are discussed in Section \ref{sec:multi-obs}. In Section \ref{sec:spectral_models}, we give a brief description of the X-ray torus models and mid-IR models we have used. The results and analysis from the X-ray spectral fitting and \xci\ SED fitting are presented in Section \ref{sec:results}. Finally, in Section \ref{sec:conclusion}, we summarize our analysis and discuss the conclusions of this paper. All reported error ranges from X-ray spectral analysis are at the $90\%$ confidence level unless stated otherwise. Through the rest of the work, we assume a flat $\Lambda$CDM cosmology with H$_0$ = 69.6\,km\,s$^{-1}$\,Mpc$^{-1}$, $\Omega_m$=0.29, and $\Omega_\Lambda$=0.71 \citep{bennett14}.

\section{Multi-wavelength observations}\label{sec:multi-obs}

NGC 6300 has been observed ten times from 2007 to 2020, using X-ray telescopes, as shown in Table \ref{Table:x-ray obs}.\footnote{It was also observed by \XMM\ in March 2001. However, we decided to exclude that observation because the data is outdated and contains corrupted auxiliary files, which cannot be used using standard data reduction processes with the Science Analysis Software (SAS).} It has also been observed multiple times in the optical-FIR band. In this section, we discuss the data reduction and data processing techniques of the different X-ray telescopes whose archival data we are using in this work. Also, we discuss the photometry extraction procedure from optical-FIR images. 

\subsection{X-ray observations and data reduction}\label{sec:xray_obs and data red}

\begin{table*}
\renewcommand*{\arraystretch}{1.2}
\centering
\caption{X-ray observational log of NGC 6300}
\vspace{.1cm}
   \begin{tabular}{ccccccc}
       \hline
       \hline       
       Instrument&ObsID&Start Time&Exposure time&Net spectral counts\tablefootmark{a}\\
       &&(UTC)&(ks)&\\
    \hline
    \suz&702049010&2007-10-17&82.6&21213\\
    \cha&10289&2009-06-03&10.2&2714\\
    \cha&10290&2009-06-07&9.8&3524\\
    \cha&10291&2009-06-09&10.2&2796\\
    \cha&10292&2009-06-10&10.2&3404\\
    \cha&10293&2009-06-14&10.2&3124\\
    \NuSTAR&60061277002&2013-02-25&17.7&6294,6457\\
    \NuSTAR&60261001002&2016-01-24&20.4&6405,6555\\
    \NuSTAR&60261001004&2016-08-24&23.5&8604,8296\\
    \cha&23223&2020-04-26&10.1&1011\\
    \hline
   \end{tabular}

   \tablefoottext{a}{The reported net spectral counts are background-subtracted total source counts. For \NuSTAR\ net counts are those of the FPMA and FPMB modules for a radius of 30\arcsec\ between 3 and 50 keV, respectively. The reported \cha\ net counts are from the ACIS-S detector, except for ObsID 23223 (ACIS-I), for a radius of 5\arcsec\ in the 0.8–7 keV energy range . The \suz\ net counts are from XIS-1 detector for a circular region of radius 150\arcsec\ in the 0.8-8.0 keV energy range.}
   \label{Table:x-ray obs}
\end{table*}
   
\subsubsection{\NuSTAR\ data reduction}

The source has been observed by \NuSTAR\ three times. The collected data have been processed using the \NuSTAR\ Data Analysis Software (NUSTARDAS) version 2.1.2. Calibration of the raw event files are performed using the \textit{nupipeline} script and the response file from \NuSTAR\ Calibration Database (CALDB) version 20211020. We utilized both focal plane modules (FPMA and FPMB) of the \NuSTAR\/. The source and background spectra are extracted from $30''$ ($\approx50\%$ of the encircled energy fraction--EEF at 10 keV) and $50''$ circular regions, respectively. The \textit{nuproducts} scripts are used to generate the source and background spectra files, along with response matrix files (RMF) and ancillary response files (ARF). Finally, using \textit{grppha}, the \NuSTAR\ spectra are grouped with at least 20 counts per bin in order to use the $\chi^2$ statistics. We have used all the three available \NuSTAR\ observational data taken from 2013 to 2016, in order to check for variability and improve the statistics of the spectra between 3 and 50 keV.  

\subsubsection{\cha\ data reduction}

NGC 6300 has been observed by \cha\ five times in 2009 and one time in 2020, using the Advanced CCD Imaging Spectrometer (ACIS). All the observations of 2009 were carried out in FAINT mode, while the 2020 observation was instead taken in VFAINT\footnote{\url{https://cxc.cfa.harvard.edu/ciao/why/aciscleanvf.html}} mode. We processed and reduced the data with the Chandra Interactive Analysis of Observations (CIAO) software version-- 4.13 and \cha\ CALDB version 4.9.5. We process the level-2 event files for each observation using the CIAO script \textit{chandra\_repro}. The source and background spectra are extracted from $5''$ (includes $> 99\%$ of EEF) and $15''$ circular regions, respectively, using the \textit{dmextract} and \textit{specextract} tools at 0.3-7.0 keV energy range. The extracted spectra are grouped using a minimum of 20 counts per bin.

\subsubsection{\suz\ data reduction}
For this work, we used a \suz\ observation taken on 2007-10-17. The data were extracted following the ABC guide\footnote{\url{https://heasarc.gsfc.nasa.gov/docs/suzaku/analysis/abc/}} from HEASARC. Running the \texttt{aepipeline}, we extracted the spectra from both the frontside (XI0, XI3) and back-side (XI1) illuminated chips unit of the X-ray Imaging Spectrometers (XIS) on a source region of 150". The response, ancillary and background files were generated running the tasks \texttt{xisrmfgen}, \texttt{xissimarfgen} and \texttt{xisnxbgen}, respectively. We then grouped the data to a minimum of 50 counts per bin. 

\subsection{Optical-FIR observations and photometry}
\label{sec:optical-FIR photometry extraction}
In order to comprehensively assess the flux of NGC 6300 over a range of wavelengths, we conducted aperture photometry using a fixed circular aperture with a radius of 9". This choice was deliberate, as it ensured the inclusion of the Full Width at Half Maximum (FWHM) of the Point Spread Function (PSF) for each filter employed in our analysis.

For the optical bands (450W, 606W, and 814W), we leveraged the highest-quality images available from the Hubble Space Telescope (HST), sourced from the Mikulski Archive for Space Telescopes\footnote{\url{https://mast.stsci.edu/}}. Expanding our measurements into the near-infrared (NIR) to far-infrared (FIR) bands, we incorporated data from the JHKs bands (Two Micron All Sky Survey - 2MASS), Spitzer IRAC and WISE, Spitzer MIPS 24 and 70 microns, Herschel PACS at 70 and 160 microns, and Herschel SPIRE at 250 microns. All data from these bands were obtained from calibrated images available in the Dustpedia database\footnote{\url{http://dustpedia.astro.noa.gr/Data}}.

For the background subtraction, we implemented a two-dimensional modeling approach for background calculation using Photutils library version 1.9\footnote{\url{https://photutils.readthedocs.io/}}. This method involved sigma clipping, a statistical technique that identifies and eliminates outliers from the dataset, while also applying a mask derived from the larger isophote provided by the Dustpedia database to exclude the galaxy as much as possible. In addition to these general background subtraction techniques, we used the star subtraction algorithm from \citet{Clark18-dustpedia} to eliminate the influence of two stars aligned with the extended part of the galaxy along the line of sight in the optical bands. Furthermore, we conducted aperture corrections and factored in the impact of Milky Way extinction on the observed brightness of NGC 6300. These procedures improve the quality of our data, and also remove any significant contamination from any other sources.


\section{Spectral modeling}\label{sec:spectral_models}

\begin{figure*}[h]
\centering
\includegraphics[width = \linewidth, trim = 0 200 0 100, clip]{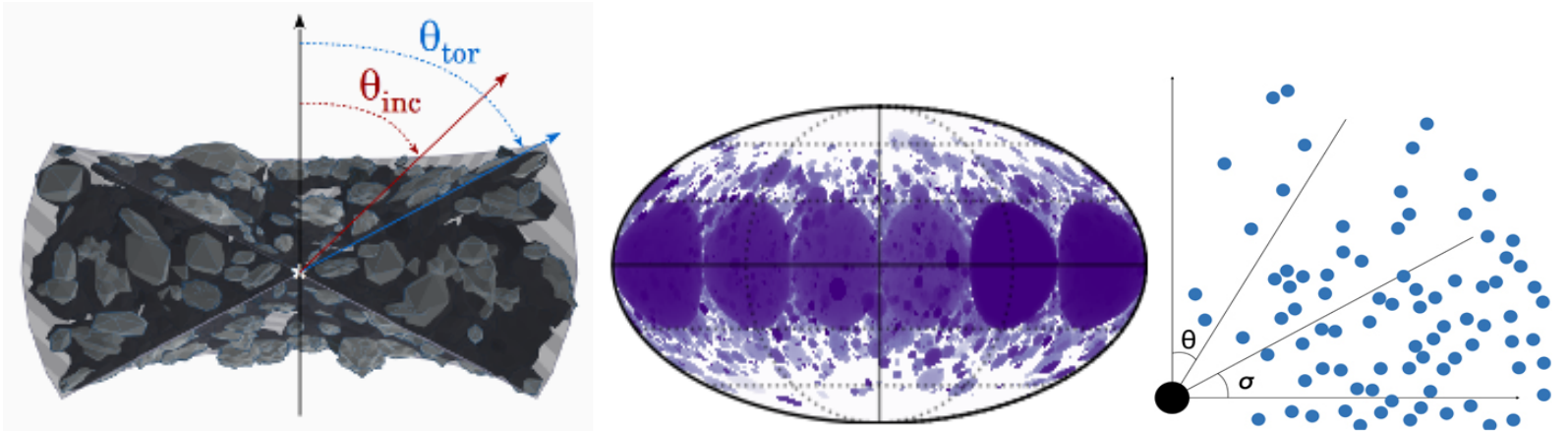} 
\caption{The physically motivated X-ray torus models, used in this work. \textit{Left:} The image shows the cross-section view of the \borus\ geometry, adopted from \citealt{balokovic2018new}, featuring a uniform spherical shape with biconical cuts along the poles. This can be made into an approximation of a clumpy medium by decoupling the LOS column density and the average column density derived from the reflection medium. The inclination angle ($\theta_{inc}$ or $\theta_i$) and the covering factor (derived from $\theta_{tor}$) is measured from the vertical axis. \textit{Middle:} The \uxclumpy\ model, from \citealt{buchner2019x}, consists of cloud clumps dispersed within a spherical geometry, including a component of Compton-thick clouds along the equatorial plane. The image shows a 2D projection of the clouds onto the sky. \textit{Right:} The  cross-section view of the \xclumpy\ model, assuming spherical clumps distributed according to a power-law along the radial direction and a Gaussian distribution along the vertical axis. This figure is adopted from \citealt{Tanimoto2019_Xclumpy}, showing $\theta$ as the inclination angle measured from the vertical axis and $\sigma$ as the torus angular width measured from the horizontal axis.}
\label{fig:torus_models}

\end{figure*}

For the X-ray spectral fitting of NGC 6300, we have used \XSPEC\ \citep{arnaud1996xspec} version 12.13.0 within the HEASOFT software (version 6.31). The metallicity is fixed at solar values from \cite{Anders_Grevesse1989}, and the photoelectric cross sections for all absorption components are determined using the method described in \cite{Verner1996}. The Galactic absorption column density is fixed at $8.01\times10^{20}$ cm$^{-2}$, following \cite{Kalberla2005Gal}. We used $\chi^2$ statistics to fit the X-ray spectra.

\subsection{Soft X-ray Model}

 Due to the large extraction region of \suz\/, we needed to handle the influence of a complex multiphase medium below 2 keV. To tackle this issue, we introduced the following soft excess model, following \cite{Torres_Alba2018}, in an attempt to produce a good fit in the soft X-ray part of the X-ray spectra from \suz\/:

\begin{equation}
\label{eq:soft_model}
\begin{aligned}
\mathrm{Soft~Model} = vapec_1 + zphabs * vapec_2 
\end{aligned}
\end{equation}

For NGC 6300, we used the variant$-apec$ or $vapec$ parameter to adjust the metal abundance pattern of the host galaxy. The first component is a standard thermal emission component and the second component is multiplied with a photoelectric absorption component $zphabs$ to represent a medium closer to the nucleus. We find the metallicity abundance ratios of a typical type II supernova explosion (SNe) properly reproduce a good fit in the soft X-ray emission part, as is expected of a medium with abundant and recent star formation. The ratios we used: (Mg, Si)/O = 1, (Ne, S)/O = 0.67, (Ar, Ca, Ni)/O = 0.46 and Fe/O = 0.27 \citep{Dupke2001_SN_metal_abundance,Iwasawa2011_metal_ratio}. The model assumes that $T_1 < T_2$, because the multiphase medium is interpreted as a combination of an outer colder unobscured medium with an inner hotter self-obscured medium. The fact that we see the opposite (which is observed within a minor population of the sample studied in \citealt{Torres_Alba2018}) for NGC 6300 (see temperatures $T_1$ and $T_2$ in Table \ref{table:NGC6300_fitting}) may mean that the $T_2$ medium may not be closer to the nucleus, but instead these media are just two distinct star-forming regions, with different properties. This model is not sufficient enough to understand all the complexities within the multiphase media of the host galaxy. As this model produces a better fit and our work is focused on characterizing the torus model, which comes from the reflection and line component ($> 2$ keV), we keep this model to fit the soft part of the spectra. 


\subsection{X-ray torus models}
\label{sec: X-ray models}

We have adopted a standard approach for analyzing the X-ray spectra of a heavily obscured AGN. This approach employs self-consistent and physically motivated smooth (uniform distribution of gas) and clumpy X-ray torus models, utilizing Monte Carlo simulations. For smooth geometry we used \borus\ \citep{balokovic2018new}, and for clumpy geometry, we used \uxclumpy\ \citep{buchner2019x} and \xclumpy\/ \citep{Tanimoto2019_Xclumpy}. Figure \ref{fig:torus_models} illustrates the physical geometry of these models, as adopted from the papers where they were originally presented. In the following sections, we provide an overview of how these models were applied in our analysis.

\subsubsection{\borus}\label{sec:borus}

The obscuring medium in \borus\ consists of a spherical geometry with biconical (polar) cut-out regions. This model is composed of three components: (a) \borus\ itself, which is a reprocessed component (including Compton-scattered $+$ fluorescent lines component), (b) $zphabs*cabs$ to include line-of-sight (LOS) photoabsorption with Compton scattering through the obscuring clouds; by this component, we multiply a $cutoffpl_1$ to account for the primary power-law continuum, and (c) finally, another $cutoffpl_2$ component is included separately, multiplied by a scaling factor $f_s<$1, to incorporate a scattered unabsorbed continuum. We approximated a clumpy medium by decoupling the components (a) and (b), since they originate from different regions. The torus covering factor ($C _{Tor}$) in \borus\ vary within the range of $0.1-1$ (i.e., the torus opening angle falls within the range of $\theta\rm _{Tor}=0\degree-84\degree$). The inclination angle $\theta\rm _{Inc}$ is kept free, ranging from $18\degree$ to $87\degree$. We used the following model configuration in \XSPEC\/:


\begin{equation}
\label{eq:borus02}
\begin{aligned}
\mathrm{Model~{\borus\/}} = C_{flux} * phabs *( borus02 + zphabs\\
* cabs * cutoffpl_1 + f_s * cutoffpl_2\\
+ \mathrm{Soft~Model}),
\end{aligned}
\end{equation}
\\

The $C_{flux}$ component is a cross calibration constant which takes into account the total flux change of different observations. We included in all the models this flux-related parameter to study any flux variability that is not related with N$\mathrm{_{H,LOS}}$. We linked all the \borus\ parameters, such as covering factor, inclination angle, N$_{\rm H,av}$ and others with each epoch, varying only N$\mathrm{_{H,LOS}}$ (from $zphabs * cabs * cutoffpl_1$) and $C_{flux}$ to study the LOS column density and flux variability, respectively. For all the models, we tied the average torus column density parameter which is derived from the reflection component, with each epoch, assuming the N$_{\rm H,av}$ doesn't change in our time scale, but only N$\mathrm{_{H,LOS}}$ changes.

\subsubsection{\uxclumpy}

The obscurer in the Unified X-ray CLUMPY (\uxclumpy\/) model has several torus geometries of interests, produced by Monte Carlo codes. This model is made up of two components: (a) $uxclumpy$ itself, which is composed of the transmitted and cold reflected component with fluorescent lines and (b) $uxclumpy\_scattered$ which takes into account the warm reflected component responsible for the scattering of the power-law from coronal emission. This model includes clumpiness and dispersion of the obscuring medium, along with an inner Compton-Thick ring of clouds modelled by \textit{CTKcover} ranging from 0 to 0.6. The cloud dispersion is modelled using the parameter \textit{TORsigma} from $6\degree$ to $90\degree$. The following model configuration is used in \XSPEC\/.

\begin{equation}
\label{eq:uxclumpy}
\begin{aligned}
\mathrm{Model~{\uxclumpy\/}} = C_{flux} * phabs *( uxclumpy + \\
 f_s * uxclumpy\_scattered + \mathrm{Soft~Model}),
\end{aligned}
\end{equation}
\\

Following the same approach we used with the \borus\ model, we linked all the \uxclumpy\ parameters in each epoch except the LOS column density (from $uxclumpy$) and $C_{flux}$ for the variability studies.

\subsubsection{\xclumpy}

The obscuring torus in the \xclumpy\ model adopted the IR CLUMPY model from \cite{Nenkova2008a,Nenkova2008b}. Each clump is assumed to be spherical with uniform gas density and radius. The clumpy distribution follows a power-law along the radial direction from inner edge to the outer edge of the torus and a Gaussian-normal distribution along the vertical axis of the torus. The model consists of four components: (a) $cabs * zphabs * zcutoffpl$ is used to compute the primary power-law emission along the LOS; (b) $f_s * zcutoffpl$ is included to reproduce the scattered unabsorbed emission; (c) $xclumpy\_reflection$ takes into account the reflected component of the torus and (d) $xclumpy\_line$ computes the fluorescence line component. All the parameters of (c) and (d) are tied with each other. In \XSPEC\ the following model configuration is used:

\begin{equation}
\label{eq:xclumpy}
\begin{aligned}
\mathrm{Model~{\xclumpy\/}} = C_{flux} * phabs *( cabs * zphabs * \\ 
zcutoffpl + f_s * zcutoffpl + xclumpy\_reflection \\ 
+ xclumpy\_line + \mathrm{Soft~Model}),
\end{aligned}
\end{equation}
\\

From the X-ray spectral fitting, we derive the hydrogen column density along the equatorial plane (N$\mathrm{_{H,eq}}$), the torus angular width ($\sigma$) within $10\degree-90\degree$ and the inclination angle ($\theta_i$) within $20\degree-87\degree$. The LOS column density is calculated directly from $cabs * zphabs$, without coupling the equatorial column density (which is calculated from the $xclumpy\_reflection$ component). We assume that the torus column density derived from the reflection component remains constant over time. Therefore, although the \xclumpy\ model typically calculates the LOS column density from the reflection component, we used the standard LOS absorption component to evaluate the column density. This method also improved the error estimates on the column density.

\subsection{Dust and mid-IR torus models}

Galaxies are composed of multiple components (e.g., gas, dust, stars, AGN) which emit radiation across all wavelengths. We used the Code Investigating GALaxy Emission (\ci\/; \citealt{Boquien2019Cigale}) included with the X-ray module of \cite{Yang2020_xcigale,Yang2022}, called `\xci\/'. It is a SED fitting code that is used to decouple the different galaxy components and study their physical properties. For NGC 6300, we have collected photometric data from the optical to FIR band at 9$''$ around the center of the galaxy (see Section \ref{sec:optical-FIR photometry extraction}). The X-ray fluxes are added from the X-ray spectral fits of \borus\/ as mentioned in Section \ref{sec: X-ray models}. Most of the results from X-ray spectral fitting using physically motivated torus models show compatible results, so we decided to use the result of only one of them. Here, we will briefly discuss on the host galaxy obscuration from stellar and dust components, but mainly focus on the torus physical properties. 

The module we used to study the star formation history (SFH) is \sfhd\/, which is a popular model that assumes a continuous star forming rate (SFR) in the galaxy. We used the stellar population library \bc\ from \cite{Bruzual2003} to compute the intrinsic stellar spectrum. The dust attenuation from UV to the NIR is computed by the \starburst\/ module based on \cite{Calzetti2000} and \cite{Leitherer2002}. Dust absorbs the optical-UV photons and re-emits at mid-IR to FIR domains covering polycyclic aromatic hydrocarbon (PAH) bands ($\sim 8 \micro\/m$), and also emission from small warm grains ($< 100 \micro\/m$) and big cold grains ($\sim 100 \micro\/m$). We modeled these dust emission processes using the \dl\/ module from \cite{Draine2014}.


\begingroup
\renewcommand*{\arraystretch}{1.2}
\begin{table*}
\centering
\vspace{.1cm}
\caption{X-ray fitting results for NGC 6300}
  \begin{tabular}{ll|ccc}
  \textbf{Parameter} && \textbf{\borus}  & \textbf{\uxclumpy\/} & \textbf{\xclumpy\/}\\
 \hline\hline
     $\chi^2$/d.o.f       &         & 2470/2571 & 2470/2571  &  2462/2562  \\
    $\chi^2_{\rm{Red}}$       &         &   0.96    &     0.96    &  0.96\\
    T$\sigma$\tablefootmark{1}        &         &   1.43$\sigma$    &     1.43$\sigma$    &  1.43$\sigma$\\
    \hline
\hline
  $kT_1$\tablefootmark{2}     &    & 0.78$_{-0.08}^{+0.08}$      & 0.78$_{-0.08}^{+0.08}$   & 0.78$_{-0.08}^{+0.08}$\\
  $kT_2$\tablefootmark{3}     &    & 0.11$_{-0.00}^{+0.00}$       & 0.12$_{-0.00}^{+0.00}$  & 0.10$_{-0.00}^{+0.00}$\\
  apec norm\tablefootmark{4} $(\times 10^{-4} )$      &  & 6.97$_{-1.13}^{+1.33}$       & 6.15$_{-1.28}^{+1.89}$   & 6.69$_{-1.07}^{+1.34}$\\
    \hline
    $\Gamma$\tablefootmark{5}     &     & 1.76$_{-0.05}^{+0.05}$       & 1.82$_{-0.06}^{+0.03}$   & 1.76$_{-0.05}^{+0.05}$\\
    N$_{\rm H,av}$\tablefootmark{6} $\times10^{24}$\,cm$^{-2}$  && 2.64$_{-0.62}^{+1.09}$       &  $\dots$   & $\dots$\\
    f$_s$\tablefootmark{7} $\times10^{-2}$       &       & 0.11$_{-0.02}^{+0.02}$ & 0.44$_{-0.14}^{+0.48}$ & 0.09$_{-0.02}^{+0.02}$\\
    C$_F$\tablefootmark{8}    &       & 0.59$_{-0.10}^{+0.12}$      & $\dots$       & $\dots$  \\
    cos$(\theta_i)$\tablefootmark{9}                &    & 0.50$_{-0.07}^{+0.10}$    & $\dots$  & $\dots$        \\
    $\theta_i$\tablefootmark{10}            &    & $\dots$  & 0$_{-*}^{+*}$     &  53.77$_{-7.55}^{+6.08}$       \\
    CTKcover\tablefootmark{11} &  & $\dots$ & 0.60$_{-0.21}^{+*}$ &  $\dots$\\
    TOR$\sigma$\tablefootmark{12} &  & $\dots$ & 24.14$_{-4.96}^{+37.33}$ & $\dots$\\
    $\sigma$\tablefootmark{13} &  & $\dots$ & $\dots$ & 28.94$_{-5.75}^{+9.46}$\\
    \hline
    $C_{\rm flux}$\tablefootmark{14} 
    & \suz\/--- 17/10/2007 & 1.09$_{-0.04}^{+0.05}$       &  1.24$_{-0.18}^{+0.21}$   & 1.13$_{-0.03}^{+0.03}$\\
    & \cha\/--- 03/06/2009 & 0.91$_{-0.06}^{+0.07}$       &  1.03$_{-0.16}^{+0.19}$   & 0.95$_{-0.06}^{+0.07}$\\
    & \cha\/--- 07/06/2009 & 1.14$_{-0.07}^{+0.08}$       &  1.29$_{-0.18}^{+0.22}$   & 1.19$_{-0.06}^{+0.07}$\\
    & \cha\/--- 09/06/2009 & 0.96$_{-0.07}^{+0.07}$       &  1.11$_{-0.18}^{+0.22}$   & 1.01$_{-0.06}^{+0.06}$\\
    & \cha\/--- 10/06/2009 & 1.09$_{-0.07}^{+0.08}$       &  1.23$_{-0.17}^{+0.21}$   & 1.14$_{-0.07}^{+0.07}$\\
    & \cha\/--- 14/06/2009 & 1.09$_{-0.07}^{+0.08}$       &  1.27$_{-0.21}^{+0.25}$   & 1.15$_{-0.07}^{+0.05}$\\
    & \NuSTAR\/--- 25/02/2013 & 1       & 1   & 1\\
    &\NuSTAR\/--- 24/01/2016 & 0.85$_{-0.03}^{+0.03}$      & 0.88$_{-0.06}^{+0.08}$   & 0.83$_{-0.04}^{+0.03}$\\
    &\NuSTAR\/--- 24/08/2016 & 0.93$_{-0.03}^{+0.03}$     & 1.00$_{-0.07}^{+0.08}$   & 0.95$_{-0.02}^{+0.02}$\\
    & \cha\/--- 26/04/2020 & 0.41$_{-0.05}^{+0.05}$       &  0.49$_{-0.11}^{+0.10}$   & 0.40$_{-0.03}^{+0.03}$\\
     \hline
    N$_{H,l.o.s.}$\tablefootmark{15}
    & \suz\/--- 17/10/2007 & 20.57$_{-0.52}^{+0.54}$       &  19.30$_{-0.57}^{+0.34}$   & 20.57$_{-0.65}^{+1.46}$\\
    & \cha\/--- 03/06/2009 & 19.60$_{-0.97}^{+1.03}$       &  18.63$_{-1.04}^{+0.71}$   & 19.71$_{-1.12}^{+1.17}$\\
    & \cha\/--- 07/06/2009 & 18.43$_{-0.76}^{+0.81}$       &  17.84$_{-1.15}^{+0.74}$   & 18.54$_{-0.86}^{+0.91}$\\
    & \cha\/--- 09/06/2009 & 20.17$_{-0.94}^{+1.00}$       &  19.21$_{-0.94}^{+0.62}$   & 20.32$_{-1.06}^{+1.11}$\\
    & \cha\/--- 10/06/2009 & 18.84$_{-0.85}^{+0.90}$       &  18.04$_{-1.04}^{+0.71}$   & 19.01$_{-0.97}^{+0.98}$\\
    & \cha\/--- 14/06/2009 & 20.54$_{-0.91}^{+0.96}$       &  19.44$_{-0.83}^{+0.56}$   & 20.94$_{-0.95}^{+1.02}$\\
    &\NuSTAR\/--- 25/02/2013 & 15.50$_{-1.08}^{+1.05}$      & 14.33$_{-1.03}^{+1.21}$   & 15.47$_{-1.29}^{+1.36}$\\
    &\NuSTAR\/--- 24/01/2016 & 16.22$_{-1.09}^{+1.09}$      & 15.26$_{-1.49}^{+1.17}$   & 15.89$_{-1.12}^{+1.36}$\\
    &\NuSTAR\/--- 24/08/2016 & 13.80$_{-0.99}^{+0.93}$      & 12.57$_{-0.82}^{+0.75}$   & 13.48$_{-0.99}^{+1.05}$\\
    & \cha\/--- 26/04/2020 & 21.75$_{-1.87}^{+2.06}$       &  20.79$_{-2.08}^{+2.43}$   & 21.96$_{-2.32}^{+2.31}$\\
   \hline
    log($flux_{2-10\rm{keV}}$)\tablefootmark{16}& & -10.38$_{-0.06}^{+0.06}$ & -10.34$_{-0.07}^{+0.07}$ & -10.41$_{-0.04}^{+0.05}$ \\
    log($flux_{10-40\rm{keV}}$)\tablefootmark{17}& & -10.29$_{-0.06}^{+0.06}$ & -10.29$_{-0.08}^{+0.07}$ & -10.29$_{-0.04}^{+0.04}$\\
    log($lum_{2-10\rm{keV}}$)\tablefootmark{18}& & 42.10$_{-0.06}^{+0.06}$ & 42.14$_{-0.07}^{+0.07}$ & 42.07$_{-0.05}^{+0.03}$\\
    log($lum_{10-40\rm{keV}}$)\tablefootmark{19}& & 42.19$_{-0.06}^{+0.06}$ & 42.19$_{-0.07}^{+0.07}$ & 42.19$_{-0.05}^{+0.03}$\\
   \hline
   \hline
\end{tabular}
  \label{table:NGC6300_fitting}
\vspace{.2cm}

\tablefoottext{1}{The Tension value for the ``true'' model.}
\tablefoottext{2,3}{\texttt{Apec} model temperature of the two mediums in units of keV.}
\tablefoottext{4}{\texttt{Apec} normalization.}
\tablefoottext{5}{\texttt{Apec} normalization.}
\tablefoottext{6}{Average hydrogen column density of the torus in units of $10^{24}$\,cm$^{-2}$.}
\tablefoottext{7}{Fraction of primary emission getting scattered instead of being absorbed, by the obscuring material.}
\tablefoottext{8}{Covering factor of the torus, from \borus\/.}
\tablefoottext{9}{Cosine of the inclination angle, from \borus\/.}
\tablefoottext{10}{Inclination angle in degree, from \uxclumpy\/ and \xclumpy\/.}
\tablefoottext{11}{Covering factor of inner thick ring of clouds, from \uxclumpy\/.}
\tablefoottext{12}{Cloud dispersion in degrees, from \uxclumpy\/.}
\tablefoottext{13}{Dispersion of torus clouds, from \xclumpy\/.}
\tablefoottext{14}{Cross normalization constant between the observations. $C_{\rm flux}$ is fixed to 1 for \NuSTAR\ observation of 2013.}
\tablefoottext{15}{LOS hydrogen column density in units of $10^{22}$\,cm$^{-2}$.}
\tablefoottext{16,17}{Average intrinsic flux within the given energy range.}
\tablefoottext{18,19}{Average intrinsic luminosity within the given energy range.}
\tablefoottext{$*$}{The upper and/or lower limits of uncertainty have reached the parameter boundary.}

\end{table*}

\endgroup



\begin{figure*}[h]
\centering
\includegraphics[width = \textwidth, trim = 0 100 0 50]{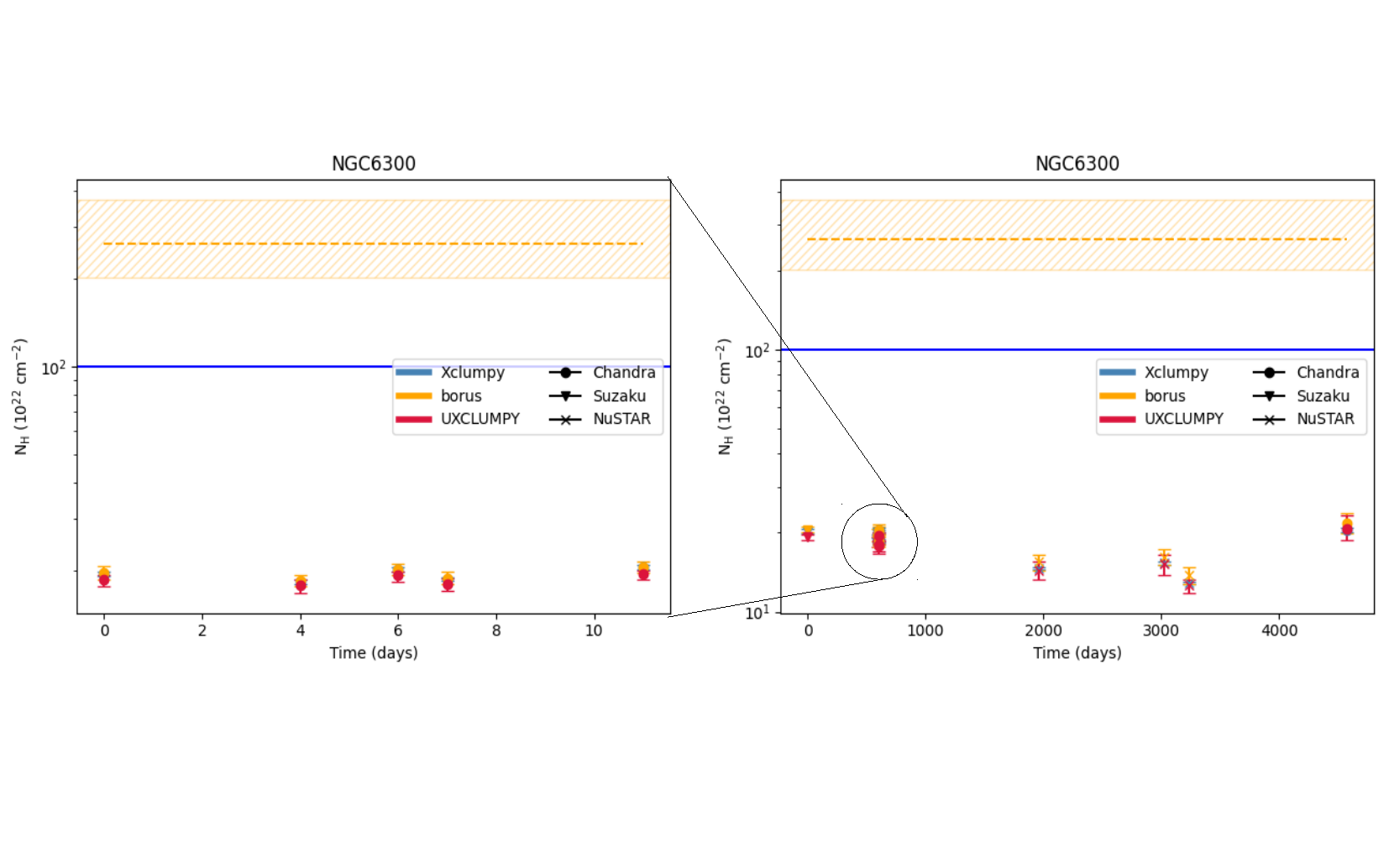} 
\caption{N$\mathrm{_{H,LOS}}$ variability in all the ten epochs from 2007 to 2020, using all the X-ray torus models. The blue horizontal line indicates the Compton-Thick column density threshold. The dashed horizontal yellow line represents the best fit value for the average $\rm{N_H}$ obtained from \borus\/. The yellow shaded area corresponds to the uncertainties associated to the average column density value. \textit{Left:} All the observations from \cha\ from 2009 are shown. \textit{Right:} All the observations, including the \cha\ ones.}
\label{fig:Nh_variability}

\end{figure*}

To model the AGN emission, we used the \skirtor\ torus model and X-ray model from \cite{Yang2020_xcigale}. Some of the input physical parameters, like the opening angle ($40\degree$), inclination angle ($50\degree, 60\degree$) and photon-index ($\Gamma = 1.8$) were selected following the best-fit values of the X-ray spectral fits. The accretion disk spectrum is set from the AGN emission module of \cite{Schartmann2005}. For the rest of the parameters, such as $\alpha_\mathrm{{ox}}$, AGN fraction, optical depth at 9.7 $\micro$m and others, we adopted a wide range of input parameters to improve the SED fitting. The X-ray fluxes are derived from the X-ray spectral fits within the range of 2-10 keV and 10-40 keV.



\section{Results and discussions}\label{sec:results}

\subsection{Results from X-ray spectral fitting}

In this section, we present the results of the X-ray fitting, as well as of the statistical analysis carried out to determine if NGC 6300 is variable, either in luminosity or in column density.

\subsubsection{Variability Evaluation}

\begingroup
\begin{table}
    \centering
    \caption{Variability analysis of NGC 6300.}
    \resizebox{\linewidth}{!}{
    \begin{tabular}{l|ccc}
  \textbf{Parameter} & \textbf{\borus\/}  & \textbf{\uxclumpy\/} & \textbf{\xclumpy\/}\\
   \hline
   \hline 
    $\chi^2$/d.o.f                & 2470/2571 & 2470/2571  &  2475/2572  \\
    T$\sigma$\tablefootmark{1}                 &   1.43$\sigma$    &     1.43$\sigma$    &  1.43$\sigma$\\
   \hline
   \hline
    $\chi^2$/d.o.f (No Var)                   & 6132/2589  & 6116/2589      &  6001/2590       \\
    T$_{\rm No\,Var}$   $\sigma$                 & 98.5$\sigma$  & 98.0$\sigma$     &  94.8$\sigma$       \\
   \hline
    $\chi^2$/d.o.f (N$_{H}$ Var)                   & 2738/2580  & 2752/2580      &  2784/2581       \\
    T$_{\rm N_H\,Var}$  $\sigma$                  & 4.4$\sigma$  & 4.8$\sigma$     &  5.6$\sigma$       \\
  \hline
    $\chi^2$/d.o.f ($C_{\rm AGN}$ Var)                   & 2631/2580  & 2670/2580      &  2657/2581       \\
    T$_{\rm C_{\rm flux}\,Var}$  $\sigma$                  & 1.4$\sigma$  & 2.5$\sigma$     &  2.1$\sigma$       \\
  \hline
    p-value N$_H$                   & 0.60  & 0.48      &  0.74     \\
   \hline
    p-value $C_{\rm AGN}$                  & 0.01  & 0.72      &  1.52E-9       \\
   \hline
   \hline
    \end{tabular}
}    
    \label{Table:tension+significance}
\end{table}
\endgroup

One of the objectives of this work is to measure the variability in the
obscuring medium (N$\mathrm{_{H,los}}$) and the variability of the intrinsic radiation coming from the central engine of NGC 6300. We used two statistical techniques to test these variabilities: Tension Statistics and Null Hypothesis. The reduced $\chi^2$ ($\chi^2_{\mathrm{Red}}$) and statistical comparisons are reported for all three models in Table \ref{Table:tension+significance}. 

A $\chi^2$ distribution is approximated as a Gaussian distribution with degrees of freedom (N). For a `true' model with perfect fit, the reduced $\chi^2$ follows a Gaussian distribution centered around the mean value of 1 and standard deviation $\sigma$ (e.g., \citealt{Andrae2010}). Following the approach outlined in \cite{Torres-alba_2023}, we used `Tension' or $T$ to define how far or close the applied model is in comparison with the `true' model fit. 

\begin{equation}
\label{eq:tension}
\begin{aligned}
T = \frac{|1-\chi^2_{\mathrm{Red}}|}{\sigma}\\
\end{aligned}
\end{equation}

Here, the standard deviation is $\sigma = \sqrt{\frac{2}{N}}$. In the first two rows of Table \ref{Table:tension+significance}, we calculated the $T\sigma$ values for the best fit of each model. The table also shows a comparison with the $T\sigma$ obtained when assuming three different scenarios: (1) no intrinsic flux or N$\mathrm{_{H,los}}$ variability between different epochs, that is, fixing all the parameters to the same value; (2) allowing only flux variability, that is, varying the fluxes for each epoch but fixing the N$\mathrm{_{H,los}}$ to a single value; (3) allowing only N$\mathrm{_{H,los}}$ variability, that is, varying the LOS column densities for each epoch but fixing the fluxes to one value. The threshold is defined as follows: when fit 1 has T$_1\sigma<3\sigma$ and fit 2 has T$_2\sigma>5\sigma$, we determine that T$_1\sigma$ is a better fit to the data than T$_2$ (\citealt{Andrae2010}). Under these assumptions, we can conclude:

--- Comparing the no variability fit (T$_{\rm No\,Var}$   $\sigma$) to the `best fit' ($T\sigma$; which includes both N$\mathrm{_{H,los}}$ and C variability), we find that we definitely require variability between different epochs to accurately describe the data, since we measure $T\sigma\sim1.4\sigma$ and T$_{\rm No\,Var}\sim100\sigma$.

--- However, for scenario (2), when we assume a condition with no N$\mathrm{_{H,los}}$ variability, the values in terms of tension are similar or close to $T\sigma$. That means, we cannot claim that N$\mathrm{_{H,los}}$ variability is required for the fit. Therefore, a pure intrinsic flux--variability scenario can, can adequately explain the data.

--- Finally, for scenario (3), when we consider a pure N$\mathrm{_{H,los}}$--variability situation, the fit is quite close to T$_{\rm C_{\rm flux}\,Var}\sigma=5\sigma$ (and above this threshold in one scenario). Thus, we can claim that some intrinsic flux variability is required to explain the data.

In summary, we observe significant variability among the epochs, but while we can confirm some flux variability is needed, the evidence for N$\mathrm{_{H,los}}$ variability is not statistically significant. Therefore, the source is definitely variable, but not purely N$\mathrm{_{H,los}}$ variable.

As an independent test to assess the source (flux and N$\mathrm{_{H,los}}$) variability, we used the p-value method. We calculated the p-value by declaring the statement of null hypothesis for obscuring column density as H$_{0N_H}$: $\mathrm{N_H}$ non-variable, and for intrinsic flux as H$_{0C_f}$: flux non-variable. This is done by computing two new $\chi^2$, using the parametric values derived from fitting the data, as follows:

\begin{equation}
\label{eq:chi2_NHlos}
\begin{aligned}
\chi^2_{N_H} = \sum_{i} \frac{(\mathrm{N_{H,LOS,i}}-\langle\mathrm{N_{H,LOS}}\rangle)^2}{\delta(\mathrm{N_{H,LOS,i})^2}}\\
\end{aligned}
\end{equation}

\begin{equation}
\label{eq:chi2_flux}
\begin{aligned}
\chi^2_{flux} = \sum_{i} \frac{(\mathrm{flux_i}-\langle\mathrm{flux}\rangle)^2}{\delta(\mathrm{flux_i})^2}\\
\end{aligned}
\end{equation}

The LOS column density for each epoch is defined as $\mathrm{N_{H,LOS,i}}$ and the average over of all the values as $\langle\mathrm{N_{H,LOS}}\rangle$. Similarly, the intrinsic flux of each epoch is classified as $\mathrm{flux_i}$ and the average of all the flux values as $\langle\mathrm{flux}\rangle$. Following the approach outlined in \cite{Barlow2002} and \cite{Torres-alba_2023}, we estimated the the asymmetric error ($\delta$) values. From the obtained $\chi^2$, we calculate the probability of the null hypothesis (p-value). We reject the null hypothesis if p-value $\leq 1\%$ for all the models, i.e., the source is variable. If it shows p-value $> 1\%$ for all the models, then we accept the null hypothesis and declare the source as non-variable for that parameter. For our source, we can conclude that it is not significantly N$\mathrm{_{H}}$ variable as all the models show p-value $> 1\%$. On the other hand, we find evidence for flux variability, as \borus\/ shows the p-value $= 0.01$ and \xclumpy\/ shows p-value $\ll 0.01$.

\begin{figure}
\centering
\includegraphics[width = 1.1\linewidth, trim = 10 0 0 0, clip]{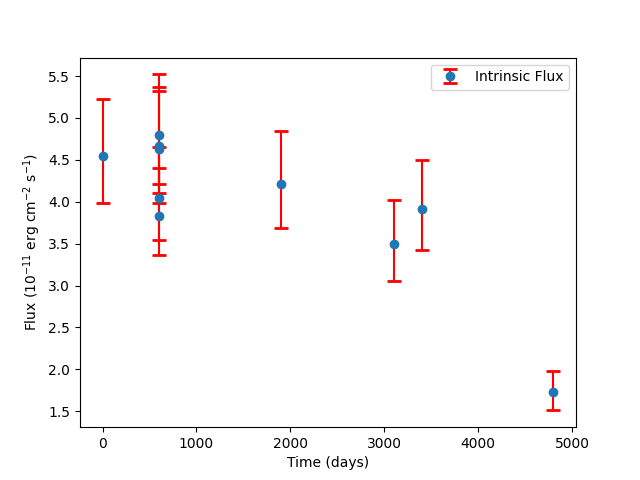} 
\caption{Variability of the intrinsic flux at 2-10 keV across all ten epochs from 2007 to 2020. A significant drop in flux is observed in the most recent \cha\ observation in 2020.}
\label{fig:intrinsic_flux}

\end{figure}

In Figure \ref{fig:Nh_variability}, we present the N$\mathrm{_{H,los}}$ variability as a function of time using all three X-ray torus models. The value of N$\mathrm{_{H,los}}$ remains similar, close to $\sim2\times10^{23}$ cm$^{-2}$ across the epochs. In Figure \ref{fig:intrinsic_flux}, we present the intrinsic flux variability as a function of time. The intrinsic flux of this source remains constant until the latest \cha\ observation in 2020, where it dropped by $40-50\%$ compared to the previous observations.


\subsubsection{Torus properties}


\begin{figure}
  \centering
  \includegraphics[width=0.5\textwidth]{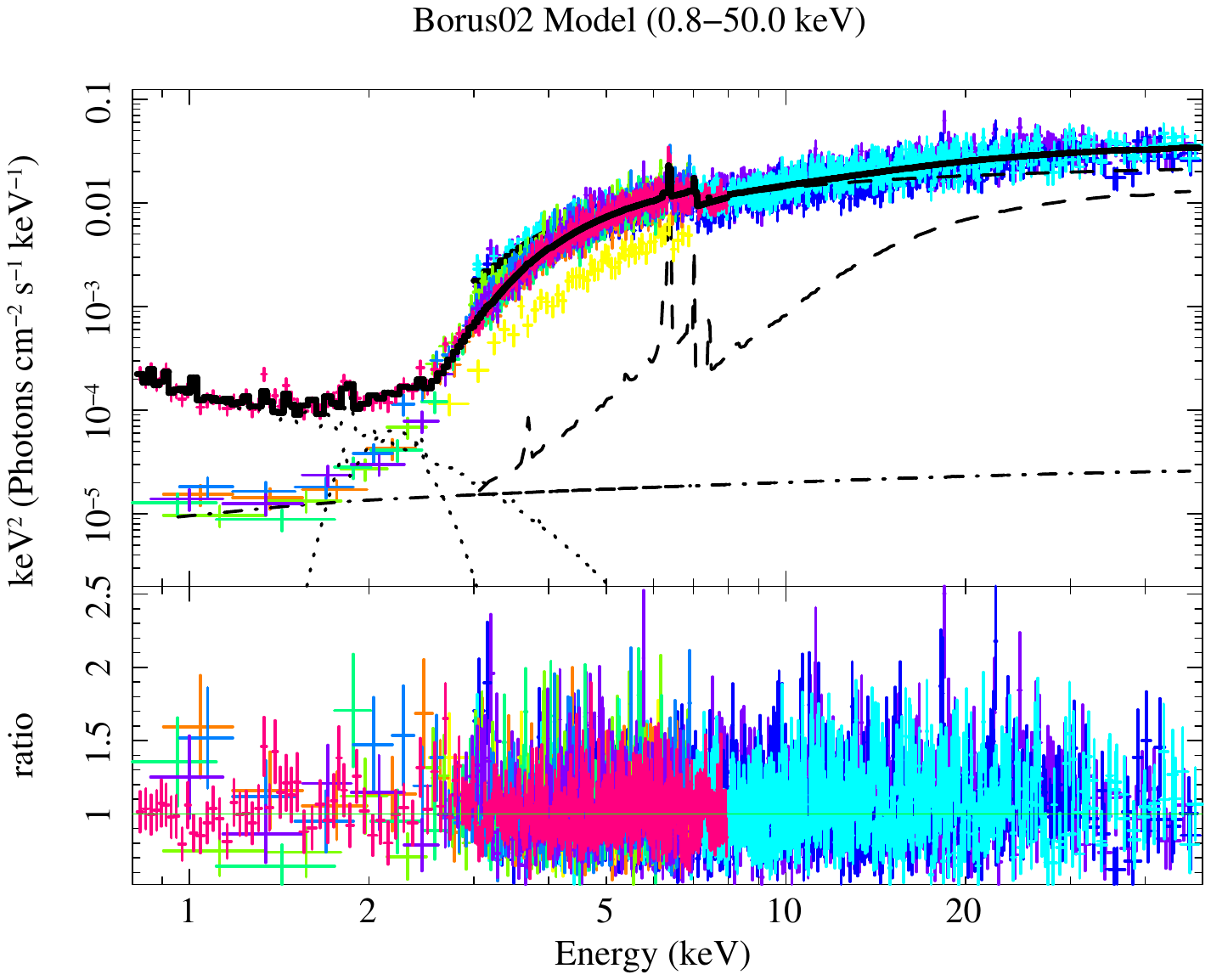}

  \includegraphics[width=0.5\textwidth]{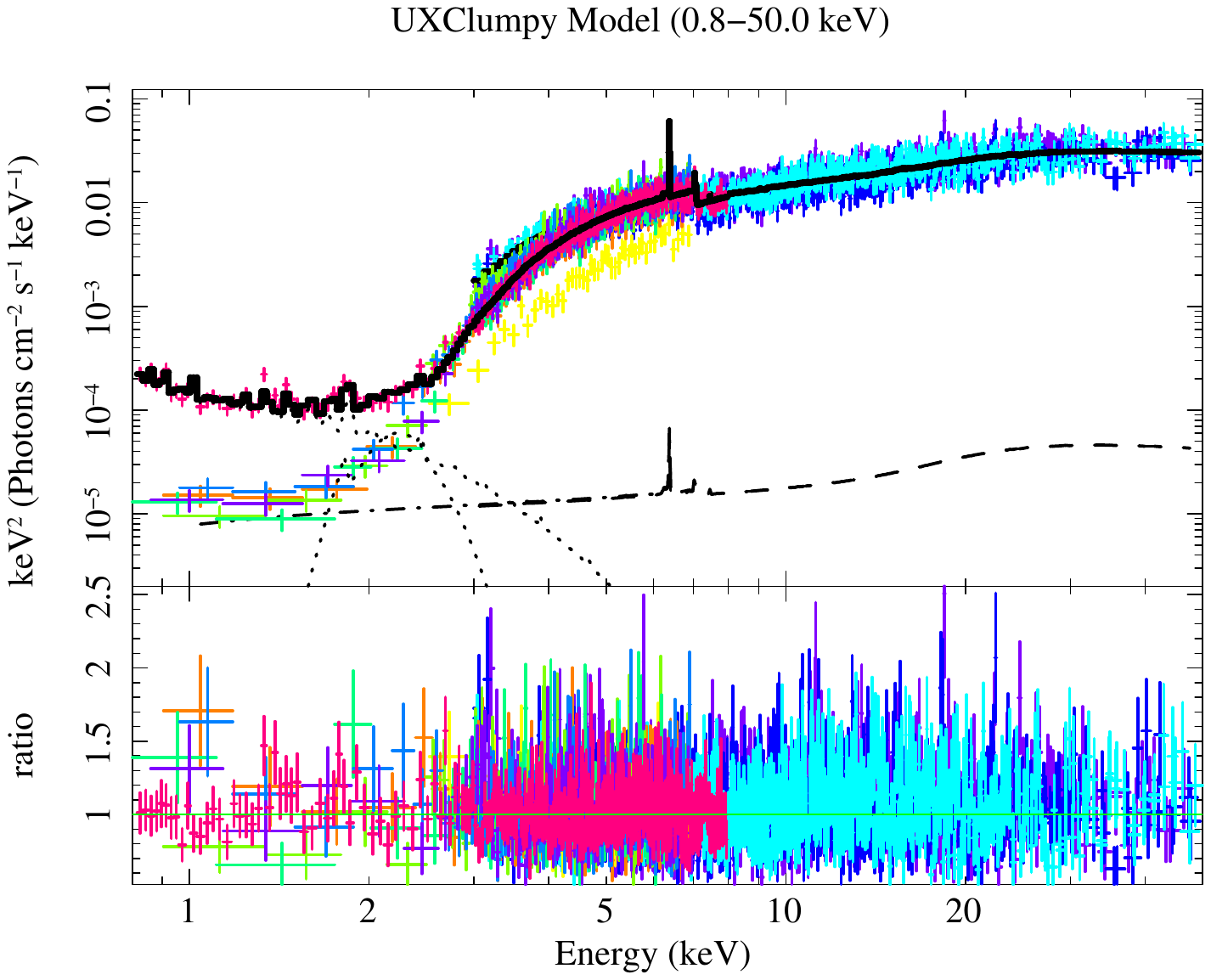}

  \includegraphics[width=0.5\textwidth]{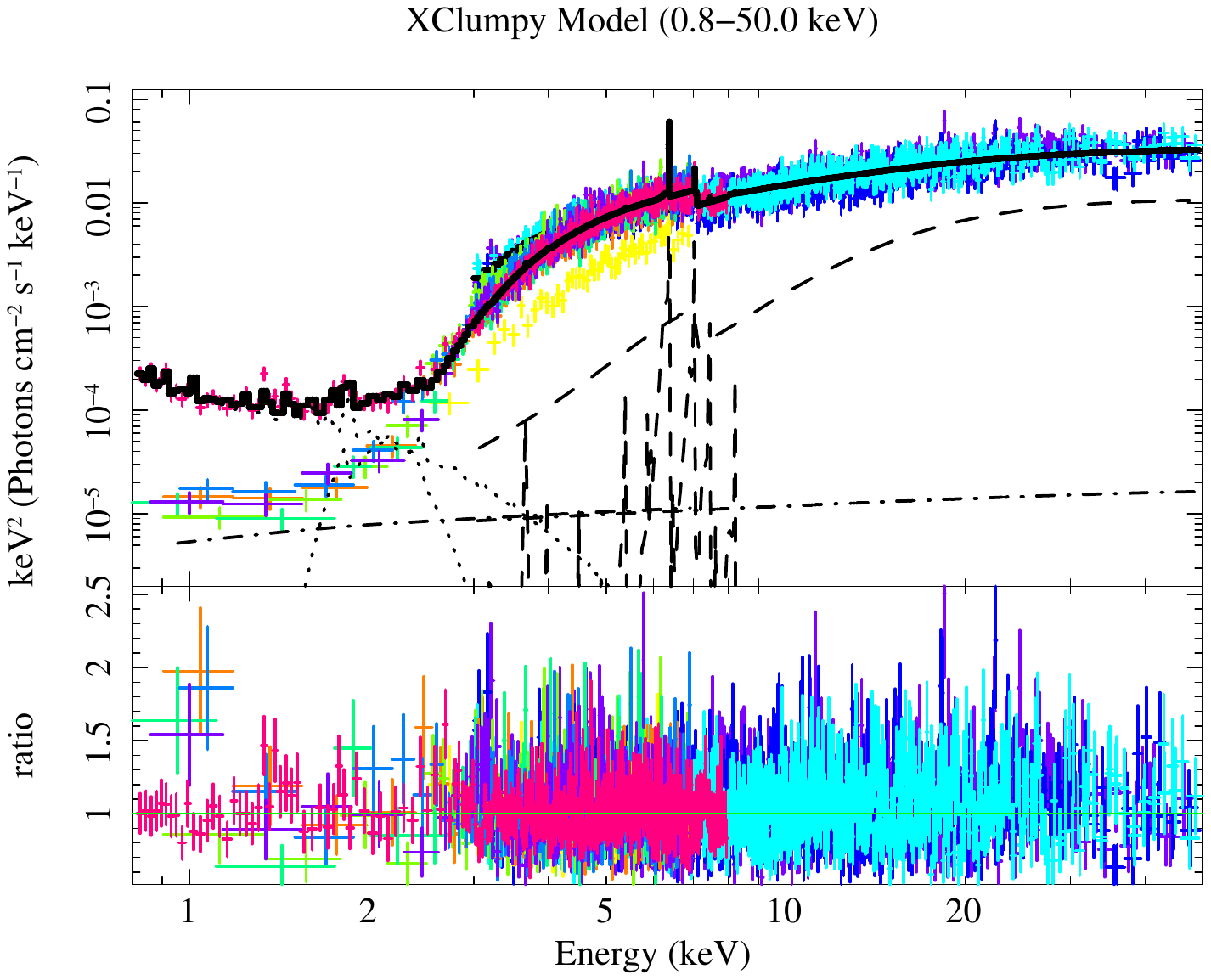}
  \caption{X-ray spectral fitting of \borus\ (top), \uxclumpy\ (middle) and \xclumpy\ (bottom) models over unfolded spectrum of NGC 6300. The \cha\ data are plotted in royal blue, orange, violet, lime, spring green and yellow. The \NuSTAR\ data are plotted in magenta, blue, cyan. The \suz\ data are plotted in crimson. The best-fit model prediction is plotted as a black solid line. The single components of the model are plotted in black with different line styles. \textit{Top}: For \borus\/, the absorbed intrinsic power-law and Compton reflection + line component is plotted with dashed line. The scattered component is marked as dot-dash line and the thermal emission from the multi-phase medium as dotted line. \textit{Middle}: In \uxclumpy\/, the Compton reflection + line component is marked as dash line and scattered continuum as dot-dash line. \textit{Bottom}: The  Compton reflection and fluorescent line component is plotted as dash lines. The scattered continuum is plotted as dot-dash line. The thermal emission from the multi-phase medium is marked as dotted line for all the three models.}
  \label{fig:borus__uxclumpy_xclumpy}
\end{figure}

\begin{figure}
\centering
\includegraphics[width = \linewidth, trim = 130 150 0 0, clip]{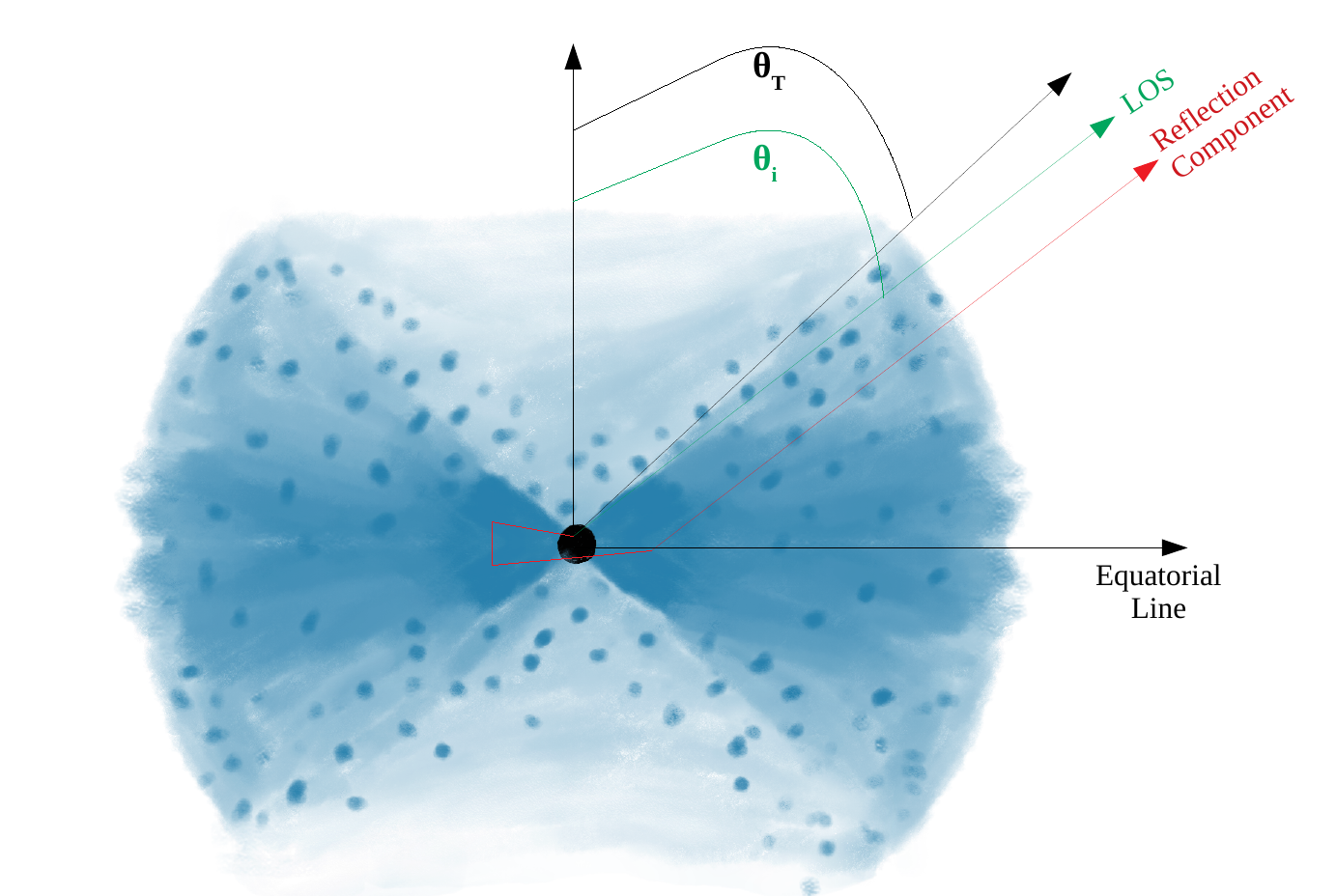} 
\caption{Schematic representation of the torus in NGC 6300, using the best-fit parameters from the physically motivated X-ray torus models. It displays the presence of Compton-Thick clouds along the equatorial region, responsible for the reflection component. The inclination angle and torus opening angle are also shown.}
\label{fig:ngc6300_diagram}

\end{figure}

 Most of the torus properties are well constrained in all the three X-ray torus models, and are consistent with each other (see Table \ref{table:NGC6300_fitting} and Figure \ref{fig:borus__uxclumpy_xclumpy}). The \borus\/ model estimates the average torus column density to be N$\mathrm{_{H,avr}} \sim 2 \times 10^{24}$ cm$^{-2}$, about one order larger than the LOS column density in each epoch, which is N$\mathrm{_{H,LOS}} \sim 2 \times 10^{23}$ cm$^{-2}$ (see Figure \ref{fig:Nh_variability}). Thus, it shows that the region mainly responsible for the reflection has significantly higher column density compared to the LOS region. Figure \ref{fig:Nh_variability} also shows how the LOS column density remains constant within the uncertainties, at all the epochs from 2007 to 2020, below the Compton-Thick threshold. It is important to note that in the previous works of \cite{Torres2021_CT-fraction,Traina2021,Sengupta2023}, the torus is classified as clumpy when $\frac{\mathrm{N_{H,avr}}}{\mathrm{N_{H,LOS}}} \neq 1$. It was interpreted as clumpy because the value of N$\mathrm{_{H,avr}}$ was calculated using one or few epochs, for each source. It was assumed that if we observe the same source in multiple epochs, the N$\mathrm{_{H,LOS}}$ values would oscillate above and below the average torus column density. However, in the recent papers of \cite{Torres-alba_2023,Pizzetti2024} and also in this work, even with more epochs, we do not observe such changes in LOS column density. Therefore, instead of assuming that it is a clumpy torus, it may be that such torus has two decoupled regions with different densities. For NGC 6300, the LOS column density remains homogeneous across all the epochs, having one order smaller column density than the Compton-Thick reflecting medium. This analysis is compatible with the results obtained using the \uxclumpy\ model, which predicts the presence of a Compton-Thick inner ring of clouds, having a high covering factor, ranging from $0.39-0.60$. The torus has moderate to high vertical dispersion (TOR$\sigma$) of the clouds. Similar to our source, the thick inner ring component was also required to model the spectra of NGC 7479 (\citealt{Pizzetti2022}), IC 4518 A (\citealt{Torres-alba_2023}) and several other sources from \cite{Pizzetti2024}. From the best-fit values of CTKcover and TOR$\sigma$, we calculated the equatorial column density to be N$\mathrm{_{H,eq}} \sim 4.4 \times 10^{25}$ cm$^{-2}$, by interpolating the N$\mathrm{_{H,eq}}$ grid within \uxclumpy\/ (\citealt{Pizzetti2024}). In comparison, the \xclumpy\/ model, which is constructed in absence of any inner thick clouds also estimates the Compton-Thick equatorial column density to be $\sim 10^{25}$ cm$^{-2}$. The inclination angle of the torus is a free parameter in \borus\/ and \xclumpy\/, and we measure it to be $\theta_i \sim 51\degree-64\degree$ within the 90\% confidence error, which is also in agreement with the results of \cite{Garcia_burillo2021} ($\sim 57\degree\/$) from ALMA observations. Figure \ref{fig:ngc6300_diagram} presents a schematic diagram of the torus, illustrating the vertical dispersion of clouds, the torus opening angle, the inclination angle, and the inner ring of clouds. 
  
To ensure the comprehensiveness of our investigation, we show that a small portion of the torus has been explored through the multi-epoch observations of the time scale of 13 years. Assuming simple Keplerian velocity for the individual clouds with independent circular orbits, the torus would have rotated by an angle of: 

\begin{equation}
\label{eq:rotation angle}
\begin{aligned}
\Delta\theta = \sqrt{\frac{\mathrm{GM}}{r^3}} \times \Delta t\\
\end{aligned}
\end{equation}

For NGC 6300, the previous papers like \citet[][from the mass-buldge luminosity correlation equation]{Khorunzhev2012} and \citet[][from molecular gas radial velocity]{Gasper2019} estimated the SMBH mass $\sim 10^7$ M$_\odot$. Assuming the outer edge of this torus is somewhere between 1 and 30 pc
(from \citealt{Gasper2019, Garcia_burillo2021, Garcia_bernete2022}), we calculate the torus have rotated between $0.001\degree-0.16\degree$. This distance, corresponds to a physical size of $\sim5 \times 10^{-4}-3 \times 10^{-3}$ pc. Therefore, the torus appears to be mostly homogeneous throughout this region, based on all the observations over the past 13 years. This suggests that we are either observing an AGN with a very uniform torus along the LOS, or that, in a rotating torus scenario, the `clumps' need to be around $5 \times 10^{-4}$ to $3 \times 10^{-3}$ pc in size. Thus, it would give the impression that we are observing through a homogeneous medium, over a period of one or two decades.


\subsection{Results from optical-FIR SED fitting}

\begin{figure}
\centering
\includegraphics[width = \linewidth, trim = 80 0 40 0, clip]{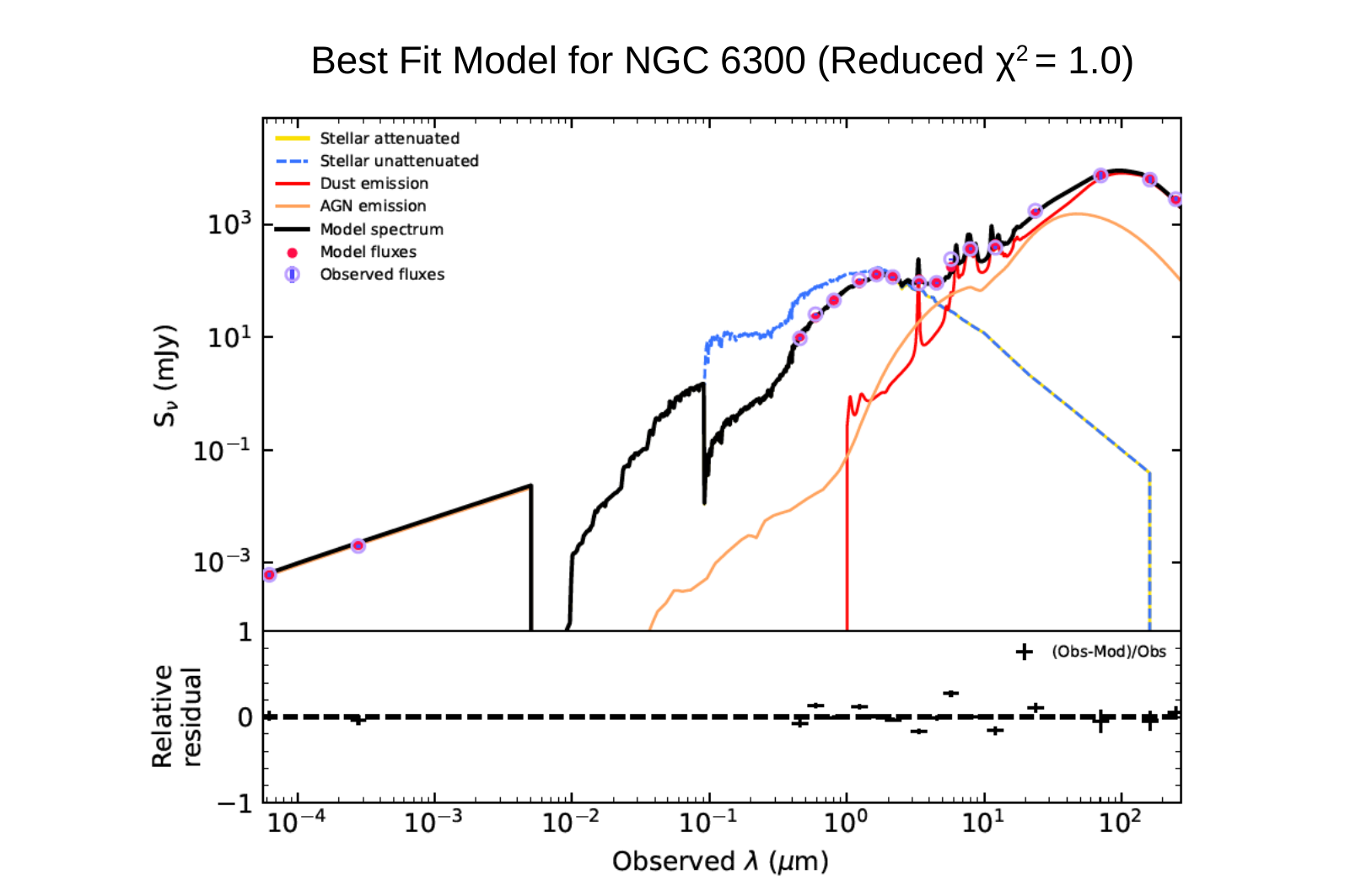} 
\caption{SED fitting of NGC 6300 using \xci\/: with AGN (orange line), host dust or stellar absorption (blue dashed line) and host dust emission (red solid line) components. The individual components are reported in the top-left part of the figure.}
\label{fig:xcigale_sed_fit}

\end{figure}

The SED-fitting result for the best model in \xci\ is shown in Figure \ref{fig:xcigale_sed_fit}. The best-fit parameters of both the AGN (disk + torus) and the stellar components are reported in Table \ref{table:xcigale}. Below we discuss the important AGN and dust properties that are responsible for the obscuration, from the SED fit.  

\subsubsection{AGN properties}

The \xci\ SED-fitting provides us with the observed AGN disk luminosity L$_\mathrm{{disk,i}} = (4.39 \pm 3.43) \times 10^{40}$ erg $\rm{s}^{-1}$. It comes out to be 100 times weaker than the intrinsic disk luminosity averaged over all directions, due to absorption along the LOS media of torus. The optical depth of the average edge-on torus at 9.7 $\micro$m is also around half of the estimated value from the SED fits on BCS sample of \cite{Garcia_bernete2022}. The fit shows optical to X-ray spectral index $\alpha_\mathrm{{ox}}=-1.25\pm0.04$, which is slightly lower than the mean value ($\sim-1.5$; \citealt{Silverman2005,Lusso2010}) observed from the deep field surveys. The ratio of the AGN luminosity with respect to the total IR luminosity, i.e., AGN fraction, is found to be around $25\%$. The AGN luminosity averaged over all the directions from the SED fitting is $\sim 4.5 \times 10^{42}$ erg s$^{-1}$, from the SED fitting. The observed AGN dust i.e., the dust in the torus and polar dust region, re-emits with the luminosity L$_\mathrm{{dust,i}} \sim 10^{42}$ erg s$^{-1}$. Using the luminosity at $B$-band (440 nm) from the SED fitting ($\sim1.42\times10^{42}$ erg s$^{-1}$) and the optical bolometric correction factor $\kappa_{O,bol}\sim5.13$ (\citealt{Duras2020}), we derive the bolometric luminosity $\rm{L_{bol}}=\kappa_{bol}\,\rm{L_{B-band}} = (7.27 \pm 0.14)\times10^{42}$ erg s$^{-1}$. In comparison, the bolometric luminosity calculated from the X-ray spectral fitting (see Table \ref{table:NGC6300_fitting}) and using the X-ray bolometric correction factor $\kappa_{X,bol}\sim15.45$ (\citealt{Duras2020}) is $\rm{L_{bol}}=\kappa_{X,bol}\,\rm{L_{2-10keV}} = (1.94 \pm 0.06) \times 10^{43}$ erg s$^{-1}$, which is $\sim2.7$ times the one derived from the optical analysis. We adopted the mean of these two derived bolometric luminosities (i.e., $\rm{L_{AGN,bol}}\sim1.33\times10^{43}$ erg s$^{-1}$), to proceed with further calculations. From the estimated SMBH mass $\sim 3.89 \times 10^7$ M$_\odot$ \citep{Khorunzhev2012} for NGC 6300, we calculate the L$_\mathrm{Edd} = 4.90 \times 10^{45}$ erg s$^{-1}$\footnote{Using the formula L$_\mathrm{Edd} = 1.26 \times 10^{38} \frac{\mathrm{M_{BH}}}{\mathrm{M_{\odot}}}$ erg s$^{-1}$.}. Thus, the Eddington ratio\footnote{$\lambda_\mathrm{{Edd}} = \frac{\mathrm{L_{AGN,bol}}}{\mathrm{L_{Edd}}}$.} comes out to be $\lambda_\mathrm{{Edd}} \sim 2 \times 10^{-3}$, which is almost one order lower than that observed in \cite{Koss2017} and BAT Complete Seyfert (BCS) sample of \cite{Garcia_bernete_2016}.

We calculate the BH accretion rate from the relation $\dot{M} = \frac{\mathrm{L_{AGN,bol}}}{\eta c^2}$ by adopting a canonical value of $\eta = 0.1$ \citep{Soltan1982}, and found to be $\dot{M} \sim 2.3\times10^{-3}$ M$_{\odot}$/yr $<< \dot{M}_{\rm{Edd}} \sim 1.2$ M$_{\odot}$/yr. It is also possible that we are not observing a classical $10\%$ efficiency (i.e., the value of $\eta$) from the accretion disk. AGN with such low accretion rate are often assumed to have radiatively inefficient accretion flow (RIAF) within the accretion disk. Such RIAF disks could produce advection dominated accretion flow (ADAF; \citealt{Yuan2014,Esin1997,narayan1994advection}) around the inner regions of the accretion disk. For ADAF cases, the gas density within the accretion medium is assumed to be lower than the standard geometrically thin accretion disk (\citealt{shakura1973reprint}), for which the energy generated within the disk gets advected inward instead of escaping the disk, forming a geometrically thick accretion disk. NGC 6300 is an obscured AGN, which falls within the range of radiatively inefficient ADAF solutions. On the other hand, many systems with CCA have $\lambda_\mathrm{{Edd}} \sim 1 \times 10^{-3}$ (\citealt{Gaspari2017}). A quiescent CCA is another likely scenario, given that the circum-nuclear media may consists of clumpy gas distribution. In that case, we are observing a quiescent period of activity, in which hot modes tend to dominate (``sunny weather''), until a new phase of precipitation triggers stronger variability and AGN feedback, e.g., via CCA (\citealt{Gaspari2013}).

From the best-fit model, we can also obtain the mid-IR luminosity at 12.3 nm, which is $\lambda\mathrm{L_\lambda} = 2.9 \times 10^{42}$ erg s$^{-1}$. Using the mid-IR vs X-ray luminosity correlation equation from equation (2) of \cite{Gandhi2009}, we derive the predicted L$_{2-10 \mathrm{keV}} \sim 2 \times 10^{42}$ erg s$^{-1}$. This value is very close to the value obtained from the X-ray spectral fit, where the displaying the intrinsic X-ray luminosity varies within the range L$_{2-10 \mathrm{keV}} \sim 1.2-1.4 \times 10^{42}$ erg s$^{-1}$. This agreement validates the fact that high resolution mid-infrared photometry can accurately proxy the intrinsic X-ray luminosity of local Seyfert galaxies like NGC 6300.



\subsubsection{Dust and stellar properties}

From Table \ref{table:xcigale}, we obtain the combined luminosity from stellar and dust component i.e. host galaxy luminosity as L$_{\mathrm{host}} = (3.54 \pm 0.14) \times 10^{43}$ erg s$^{-1}$. It shows that the stellar dusts along the LOS is almost one order more luminous ($\frac{\mathrm{L}_{\mathrm{host}}}{\mathrm{L_{AGN}}} \sim 8.8$) than the AGN (torus + polar dust), in the IR band. The SFR of NGC 6300 is found to be very low $\sim 0.19$ $\mathrm{M_{\odot} yr^{-1}}$ from the \sfhd\ module of \xci\/. We further derived the SFR value from \cite{Kennicutt1998} relation $\mathrm{log(SFR/M_{\odot} yr^{-1}) = log(L_{FIR}/ erg s^{-1}) - 43.34}$, assuming $\mathrm{log(L_{FIR}) \approx log(L_{dust})}$. The result showed SFR = $0.59 \pm 0.04$ $\mathrm{M_{\odot} yr^{-1}}$, which is compatible with the \xci\ value. The fit shows a dust mass $\sim 4.56 \times 10^{36}$ kg at radius 9" ($\sim600$ pc). In comparison, from the ALMA observation at 0.1" ($\sim3-4$ pc), the derived dust mass $\sim 6 \times 10^{35}$ kg (\citealt{Garcia_burillo2021}), showing most of the dust concentration is in the nuclear region.     



\begin{table*}
\renewcommand*{\arraystretch}{1.2}
\centering
\caption{Physical parameters of NGC 6300 from best-fit SED}
\begin{tabular}{c|cc|c}
\hline
Model Component & Parameters & Units & Values \\
\hline
\hline
 &agn.fracAGN & & 0.23 $\pm$ 0.04\\
 &agn.t & & 5.71 $\pm$ 1.43\\
 &xray.alpha\_ox & & -1.25 $\pm$ 0.04\\
AGN &xray.gam & & 1.78 $\pm$ 0.08\\
 &agn.accretion\_power & W $\times 10^{35}$ & 4.5 $\pm$ 0.9\\
 &agn.disk\_luminosity & W $\times 10^{33}$ & 4.4 $\pm$ 3.4\\
 &agn.luminosity & W $\times 10^{35}$ & 4.0 $\pm$ 0.7\\
\hline
 &dust.alpha & & 2.06 $\pm$ 0.05\\
Dust &dust.qpah & & 3.01 $\pm$ 0.69\\
and &dust.umin & & 5.29 $\pm$ 1.30\\
Stellar History &dust.mass & kg $\times 10^{36}$ & 4.6 $\pm$ 0.7\\
 &dust.luminosity & W $\times 10^{36}$ & 1.3 $\pm$ 0.1\\
 &sfh.age\_main & Myr & 4025 $\pm$ 571\\
 &sfh.sfr & M$_{\odot}$ yr$^{-1}$ & 0.19 $\pm$ 0.02\\
 &stellar.lum & W $\times 10^{36}$ & 2.25 $\pm$ 0.11\\
 &stellar.m\_gas & M$_{\odot}$ $\times 10^{9}$ & 1.69 $\pm$ 0.23\\
 &stellar.m\_star & M$_{\odot}$ $\times 10^{9}$ & 2.22 $\pm$ 0.25\\
\hline
\hline
\end{tabular}
\label{table:xcigale}
\end{table*}


\section{Summary and conclusions}\label{sec:conclusion}
We have analysed multi-epoch X-ray data of NGC 6300 from 2007 to 2020. Using physically motivated X-ray torus models, we have studied column density and flux variability of the X-ray spectra within the energy range 0.8 keV to 50.0 keV. We also estimated torus properties like inclination angle, covering factor, torus cloud dispersion, average column density and others. For a comprehensive picture of the nuclear obscuring medium, we used the X-ray results to fit optical-FIR SED over photometric data points. In this section, we summarise our conclusions: 

\begin{enumerate}

    \item NGC 6300 was reported as a changing-look AGN candidate. However in agreement with the results of the last $\sim20$ years, even with the \cha\ observation of 2020, this source does not show any N$\rm _{H,LOS}$ variability. We used both smooth and clumpy torus models, to study the statistical significance of any variability nature along the LOS column density. All the models agree that the source is non-variable in terms of N$\rm _{H,LOS}$ having value around $2\times10^{23}$ cm$^{-2}$. In conclusion, we observe the source through a Compton-Thin gas distribution.

    \item While there is no N$\rm _{H,LOS}$ variability, the observation of 2020 showed a significant flux variability in the energy band E$=0.8-7.0$ keV. The flux dropped by $\sim40-50\%$ in comparison with all the other observations, since 2007. Two of the three torus models also confirm with high statistical significance that there is an existing signature of intrinsic flux variability for this source. 

    \item The N$\rm _{H,LOS}$ values are almost homogeneous and $\sim10$ times smaller than the \borus\ calculated N$\rm _{H,avr}$, which is the average column density derived from the reflection dominated region. The model \uxclumpy\ predicts the presence of the inner CT-ring of gaseous medium is responsible for the reflection dominated spectra, having equatorial column density $\sim10^{25}$ cm$^{-2}$. \xclumpy\ also shows that, along the equatorial region, the torus gets highly over-dense ($\sim10^{25}$ cm$^{-2}$) compared to the LOS region. In our timescale, we estimated to have observed $\sim5 \times 10^{-4}-3 \times 10^{-3}$ pc angular region of the torus. 


    \item The mean bolometric luminosity is evaluated from the optical-IR SED fitting and X-ray spectral fitting. We further estimated sub-eddington accretion ($\lambda_{\rm{Edd}}\sim2\times10^{-3}$), which falls within the range of ADAF accretion flow, with geometrically thick disk.

    \item In terms of a general Black Hole Weather framework (see diagram in figure 1 of \citealt{Gaspari2020}), the observations suggest that NGC 6300 is continuing to experience a quiescent period with purely hot-mode variations (``sunny weather'') but with a micro/meso-scale clumpy structure, likely residual of the previous cooling cycle. We expect a subsequent reactivation of the AGN feedback once the macro-scale precipitation resumes, triggering a next cycle of CCA, which will be highlighted by boosted $\rm{N_H}$ variability.

    \item SED fitting on optical-IR photometry also validate the obscuring nature of torus. We find the mid-IR photometry SED fitting can accurately proxy the X-ray luminosity. The X-ray luminosity derived from the 12 $\mu$m luminosity is consistent with the X-ray luminosity observed from the X-ray spectral fitting. Further calculation shows low SFR with high dust concentration near the nuclear region.

    \item Joint X-ray and mid-IR analysis of AGN SED helps to characterize the obscuring nature of torus: IR emission of torus, optical depth, accretion rate, dust and gas influence in obscuration, stellar influence. The results are consistent with recent ALMA observations.

\end{enumerate}

\section{Acknowledgements}

 This research has made use of the data from the X-ray telescope \NuSTAR\ data, and also used the software NuSTARDAS, jointly developed by the ASI Space Science Data Center (SSDC, Italy) and the California Institute of Technology (Caltech, USA). The scientific results reported in this paper are also based on observations made by the X-ray observatories like \suz\ and \cha\/, and have used NASA/IPAC Extragalactic Database (NED), which is operated by the Jet Propulsion Laboratory, Caltech, under the contract with NASA. We acknowledge the use of the software packages HEASARC and CIAO. DS acknowledges the PhD and ``MARCO POLO UNIVERSITY PROGRAM'' funding from the Dipartimento di Fisica e Astronomia (DIFA), Università di Bologna. DS and SM acknowledge the funding from INAF ``Progetti di Ricerca di Rilevante Interesse Nazionale'' (PRIN), Bando 2019 (project: ``Piercing through the clouds: a multiwavelength study of obscured accretion in nearby supermassive black holes''). M.G. acknowledges support from the ERC Consolidator Grant \textit{BlackHoleWeather} (101086804). IEL acknowledges support from the Cassini Fellowship program at INAF-OAS and the European Union's Horizon 2020 research and innovation program under Marie Sklodowska-Curie grant agreement No. 860744 ``Big Data Applications for Black Hole Evolution Studies'' (BiD4BESt).

\bibliographystyle{aa}
\bibliography{1biblio}

\end{document}